\documentclass[twocolumn,         
               showpacs,longbibliography,            
               showkeys,preprintnumbers,     
               aps,                 
               prd,          	    
               letterpaper,             
               superscriptaddress,      
               nofootinbib,         
               tightenlines,        
               floats,floatfix      
               ]{revtex4-1}
\usepackage{physics}
\usepackage{epsf}
\usepackage{graphicx,color}
\usepackage{subfigure}
\usepackage{latexsym}
\usepackage{amsmath,amssymb}        
\usepackage[colorlinks=true,linkcolor=blue,citecolor=blue]{hyperref}
\usepackage{mathrsfs}
\usepackage{comment}
\usepackage{soul}
\usepackage{cancel}
\definecolor{purple}{rgb}{0.58,0.0,0.83}
\definecolor{orange}{rgb}{1,0.5,0}
\DeclareSymbolFontAlphabet{\mathrsfs}{rsfs}
\DeclareMathAlphabet{\mathcal}{OMS}{cmsy}{m}{n}

\begin{document}


\title{Simulation of Gaussian Wave Packets used to Illustrate Elementary Quantum Mechanics Scenarios}


\author{Francisco Guzman-Cajica}
\email{2002236d@umich.mx}
\affiliation{Facultad de Ciencias F\'isico Matem\'aticas, Universidad
              Michoacana de San Nicol\'as de Hidalgo. Edificio ALFA, Cd.
              Universitaria, 58040 Morelia, Michoac\'{a}n,
              M\'{e}xico.}  

\author{Francisco S. Guzm\'an}
\email{francisco.s.guzman@umich.mx}
\affiliation{Instituto de F\'{\i}sica y Matem\'{a}ticas, Universidad
              Michoacana de San Nicol\'as de Hidalgo. Edificio C-3, Cd.
              Universitaria, 58040 Morelia, Michoac\'{a}n,
              M\'{e}xico.}  


\date{\today}


\begin{abstract}
In this paper we numerically solve the time dependent Schr\"odinger equation for scenarios using wave packets. These examples include the free wave packet, which we use to show the difference between group and phase velocities, the packet in a harmonic oscillator potential with non-trivial initial conditions in one and two dimensions, which is compared with their classical analogs to show how Ehrenfest theorem holds. We also include simulations of the diffraction through the single and double slit potentials, the refraction with a step potential and the dispersion by a central potential. The aim of this paper is to illustrate with simulations,  nowadays easy to implement, scenarios that can help explaining the basics of the wave-particle duality.
\end{abstract}




\maketitle

\section{Introduction}
\label{sec:intro}

It is common to find illustrative examples of quantum systems in books and courses, mostly concerning stationary scenarios, like the particle in a box, the particle in a harmonic oscillator potential and the Hydrogen atom found in typical text books. Nowadays, with the help of numerical methods and affordable computing power, it is relatively simple to construct non-stationary scenarios with certain dynamics, which help illustrating some basics like the wave-particle duality, understood from solutions of Schr\"odinger equation.

With this aim we produced a number of programs that solve Schr\"odinger equation for some interesting scenarios involving modulated wave packets whose dynamics concerns the particle and wave nature described by the wave function. The cases we selected for illustration concern typical problems with nearly trivial potentials, however using non-stationary and non-trivial initial conditions. We also point to supplemental material that includes animations of the problems solved here, that serve as complement of the snapshots included in paper and may help for educational purposes in the classroom \cite{GeneralSE}. We believe that this work can be a complement of some educative analytical \cite{RMFQuantumExact} and numerical \cite{RMFSchro} solutions of Schr\"odinger equation.

We first present problems defined in one spatial dimension and start with the well studied case of the free wave packet. We use this example to show how a Gaussian packet spreads during the evolution as predicted by the exact solution, and also to point out the difference between group and phase velocities. We evolve a packet subject to the effects of the Harmonic Oscillator potential for non trivial initial conditions, first the quasi-classical state and later a rather arbitrary Gaussian wave packet to show the differences in terms of distortion of the density. We also solve the classical analog to verify that the Ehrenfest theorem holds.

Problems defined on two spatial dimensions are also solved, specifically the evolution of a packet on a two-dimensional harmonic trap, and it is shown how the Ehrenfest theorem holds for a system with two degrees of freedom. In more elaborate scenarios the diffraction of the wave-packet by a single and double slit potentials are studied, for which we calculate the interference pattern. We also present the reflection and refraction by a step potential. This case  concerns the deflection of the wave front, and implies the Snell law for the wave packet evolving according to Schr\"odinger equation. Finally we study the dispersion by a central potential, which according to theory \cite{cohen} a modulated plane wave interacting with a central potential would produce a spherical wave front.

The paper is organized as follows. We first specify the numerical methods used to solve Schr\"odinger equation in Section \ref{sec:nm}, then in Sections \ref{sec:p1d} and \ref{sec:p2d} we describe the problems in one and two spatial dimensions. Finally in Section \ref{sec:comments} we present some final comments.

\section{Numerical methods}
\label{sec:nm}

The fully time-dependent Schr\"odinger equation for a particle of mass $m$ subject to the action of a potential $U$ reads

\begin{equation}
    {\rm i} \hbar\frac{\partial \Psi}{\partial \tau} = -\frac{\hbar^2}{2m}\nabla_{\xi}^2 \Psi + U(\vec{\xi},\tau)\Psi
\end{equation}

\noindent where $\Psi=\Psi(\vec{\xi},\tau)$ and ${\rm i} := \sqrt{-1}$; we define this number because, below we will keep the traditional notation of $i$ for integer labels of the numerical domain and $\hat{\textbf{\i}}$ for the unitary basis vector, both along the $x-$direction. In order to solve this equation numerically, we rescale variables according to the transformation:

\begin{eqnarray}
\vec{x} = \frac{\vec{\xi}}{a_0}&,&t = \frac{E_0}{\hbar}\tau, \nonumber \\
V(\vec{x},t) &=& \frac{1}{E_0}U(\vec{\xi},\tau), \nonumber\\
\psi(\vec{x},t) &=& \sqrt{a_0^k}\Psi(\vec{\xi},\tau),\nonumber\\
\hat{p} &=& \frac{a_0}{\hbar}\hat{p_{\xi}} = -i\nabla_{x}, \nonumber\\
a_0 &=& \text{Bohr radius}, \nonumber\\
E_0 &=& \frac{\hbar^2}{ma_0^2},
\end{eqnarray}

\noindent where $a_0$ is a length scale appropriate for each problem to be treated. If we are working with atomic scales, it could be one Angstrom or Bohr's radius. These new variables are dimensionless and will simplify numerical calculations. In the redefinitions above, the wave function is also rescaled to ensure it has norm one, while $k$ is the dimension of the spatial domain, in this paper one and two. The resulting equation reads

\begin{equation}
    {\rm i}\frac{\partial \psi}{\partial t} = -\frac{1}{2}\nabla_{x}^2 \psi + V(\vec{x},t)\psi,
\label{eq:SchroedingerCodeUnits}
\end{equation}

\noindent where $\psi = \psi(\vec{x},t)$. We call these dimensionless units {\it code units} and they will be used from this point on.

Equation (\ref{eq:SchroedingerCodeUnits}) is solved numerically as an initial value problem provided some initial conditions for $\psi$. The numerical method used for integration is based on Finite Differences defined on a uniformly discrete domain for one and two-dimensional scenarios.

{\it One-dimensional problems.}. The space-time domain of solution is $D=[x_{min},x_{max}]\times[0,t_f]$. The numerical  domain is the set of points $(x_i,t^n)\in D$ such that $x_i=x_{min}+i\Delta x$, where $\Delta x = (x_{max}-x_{min})/N_x$ and $i=0,1,...,N_x$ where $N_x$ is the number of cells along the spatial domain, whereas $t^n=n\Delta t$ with $\Delta t=CFL \Delta x^2$ and $n=0,1,...,N_t=t_f/\Delta t$, where $CFL$ stands for the Courant-Friedrichs-Lewy factor between spatial and time resolutions. In this domain the wave function and potential at the arbitrary point $(x_i,t^n)$ are denoted by $\psi^n_i$ and $V^n_i$ respectively.

For the evolution from time $t^n$ to $t^{n+1}$ we implement the Crank-Nicolson average, which leads to the following second order accurate evolution scheme for Eq.  (\ref{eq:SchroedingerCodeUnits}):

\begin{eqnarray}
(-\alpha)\psi^{n+1}_{i-1} &+& (1+2\alpha+\beta V^{n+1}_i)\psi^{n+1}_{i} + (-\alpha)\psi^{n+1}_{i+1}\label{eq:1dtridiag}\\ &=&
(\alpha)\psi^n_{i-1} + (1-2\alpha-\beta V^n_i)\psi^n_{i}+(\alpha)\psi^n_{i+1},
\nonumber
\end{eqnarray}

\noindent where $\alpha=\frac{1}{4}{\rm i}\frac{\Delta t}{\Delta x^2}$, $\beta=\frac{1}{2}{\rm i}\Delta t$. This scheme defines a tridiagonal linear system that is valid for inner points $i=1,...,N_x-1$, whereas boundary conditions at $x_{min}=x_0$ and $x_{max}=x_{N_x}$ determine the equation for space labels $i=0$ and $i=N_x$. In all the examples we use a big enough domain as to impose the zero boundary condition $\psi(\partial D,t)=0$, which according to \cite{fsguzman} implies the equations 
$(1+2\alpha + \beta V^{n+1}_{0})\psi^{n+1}_0=(1-2\alpha-\beta V^n_0)\psi^n_0$ 
and 
$(1+2\alpha + \beta V^{n+1}_{N_x})\psi^{n+1}_{N_x}=(1-2\alpha-\beta V^n_{N_x})\psi^n_{N_x}$ 
at points with labels $i=0$ and $i=N_x$ respectively. These two equations complete the system for all $i=0,...,N_x$ and is then solved for $\psi^{n+1}_{i}$ using forward and backward substitution following the recipe in \cite{nr}.

{\it Two-dimensional problems.} In this case the domain of integration is $D=[x_{min},x_{max}]\times[y_{min},y_{max}]\times[0,t_f]$ and the numerical domain is defined by the points $(x_i,y_j,t^n)\in D$ such that $x_i=x_{min}+i\Delta x$, $y_j=y_{min}+j\Delta y$ with $\Delta x=(x_{max}-x_{min})/N_x$, $\Delta y=(y_{max}-y_{min})/N_y$ with $N_x$ and $N_y$ the number of cells along $x$ and $y$ directions, $i=0,1,...,N_x$, $j=0,...,N_y$. In all the examples we use $\Delta x=\Delta y$, so that time discretization can be defined by $t^n=n\Delta t$ with $\Delta t= CFL\Delta x^2$ and $n=0,1,...,N_t$. In this domain at point $(x_i,y_j,t^n)$ any involved function $f$ is denoted by $f^n_{i,j}$.

The evolution is carried out using the Crank-Nicolson method as well, along with the Alternating Direction Implicit (ADI) order of integration across spatial directions. With this scheme the evolution from $t^n$ to $t^{n+1}$ is implemented in three steps as follows \cite{fsguzman}:

\begin{eqnarray}
\left( 1-\frac{\rm i}{4} \Delta t ~ \delta^2_x \right) R_{i,j} &=& 
\left( 1+\frac{\rm i}{4} \Delta t ~ \delta^2_x\right) \psi^{n}_{i,j},\nonumber\\
\left( 1-\frac{\rm i}{4} \Delta t ~ \delta^2_y\right) S_{i,j} &=& 
\left( 1+\frac{\rm i}{4} \Delta t ~ \delta^2_y \right) R_{i,j},\nonumber\\
\left( 1+\frac{\rm i}{2} \Delta t ~V_{i,j}\right) \psi^{n+1}_{i,j} &=& 
\left( 1-\frac{\rm i}{2} \Delta t ~V_{i,j}\right) S_{i,j},\label{eq:ADISchro3D}
\end{eqnarray}

\noindent where $R_{i,j}$ and $S_{i,j}$ are auxiliary grid functions that update the values of the wave function after applying the derivative operator along each spatial direction $ \delta^2_x, ~ \delta^2_y$, which are the second order derivative discrete operators. In two dimensional problems we only deal with time-independent potentials, thus we omit the superindex on $V_{i,j}$ in Eq. (\ref{eq:ADISchro3D}). These are two tridiagonal and one diagonal systems of linear equations that are formulated likewise in (\ref{eq:1dtridiag}), explicitly 

\begin{eqnarray}
(-\alpha)R_{i-1,j} &+& (1+2\alpha)R_{i,j} + (-\alpha)R_{i+1,j} =\nonumber\\
&&(\alpha)\psi^{n}_{i-1,j}+(1-2\alpha)\psi^{n}_{i,j} + (\alpha)\psi^{n}_{i+1,j},\nonumber\\
(-\alpha)S_{i,j-1} &+& (1+2\alpha)S_{i,j} + (-\alpha)S_{i,j+1} =\nonumber\\
&&(\alpha)R_{i,j-1}+(1-2\alpha)R_{i,j} + (\alpha)R_{i,j+1},\nonumber\\
\psi^{n+1}_{i,j} &=& \frac{1- \beta  V_{i,j}}{1+ \beta V_{i,j}}S_{i,j},\label{eq:ADISchro3Dtrid}
\end{eqnarray}

\noindent where $\alpha=\frac{1}{4}{\rm i}\frac{\Delta t}{\Delta x^2}$ and $\beta=\frac{1}{2}{\rm i}\Delta t$. These systems are completed with equations at boundary sides by imposing  the boundary condition $\psi(\partial D,t)=0$. The boundary conditions on the numerical domain are analogous to those for the one-dimensional for $i,j=0$ and for $i=N_x, ~j=N_y$. Once with the complete systems of equations, we solve the two tridiagonal systems following the implementation in \cite{nr} of the forward-backward substitution, and the diagonal system is solved using direct substitution.\\


\section{Problems in 1D}
\label{sec:p1d}

Beyond stationary problems, time-dependent cases in one spatial dimension are the first scenarios discussed in Quantum Mechanics. Here we elaborate and discuss the simplest ones from a simulation based approach. 

\subsection{Free Gaussian Wave Packet}

Moving free particles in Quantum Mechanics are some times represented by wave packets. Most of the time, the momentum of a particle is not known with complete certainty, it does not have a definite wavelength. Their wave functions are rather a sum of waves which cover a spectrum of wave numbers. A simple way to represent this is through a Gaussian wave packet, which is nothing more than a sum of plane waves with amplitudes that vary according to a Gaussian distribution. If the wave number of maximum amplitude corresponds to $k_0$ and the standard deviation of the distribution is $\sqrt{2}/a$, then the normalized wave function of such particle at $t=0$ is \cite{cohen}:

\begin{eqnarray}
\psi(x,0) &=& \frac{\sqrt{a}}{(2\pi)^{3/4}}\int_{-\infty}^{\infty}e^{-\frac{a^2}{4}(k-k_0)^2}e^{ikx}dk \nonumber\\
&=& \left( \frac{2}{\pi a^2} \right)^{1/4}e^{ik_0x}e^{-x^2/a^2}.
\label{eq:IDFreeParticle}
\end{eqnarray}

The time evolution of such packet can be found by summing each plane wave $e^{ikx}$ multiplied by $e^{-iw(k)t}$, its time evolution. After doing the integral, the probability density of the particle happens to be \cite{cohen}:

\begin{equation}
|\psi(x,t)|^2 = \sqrt{\frac{2}{\pi a^2}}\frac{1}{\sqrt{1+\frac{4t^2}{a^4}}}\exp{ -\frac{2(x-k_0t)^2}{a^2+\frac{4t^2}{a^2}} }.
\label{eq:ExactFreeParticle}
\end{equation}

\begin{figure}
\centering
\includegraphics[width=9cm]{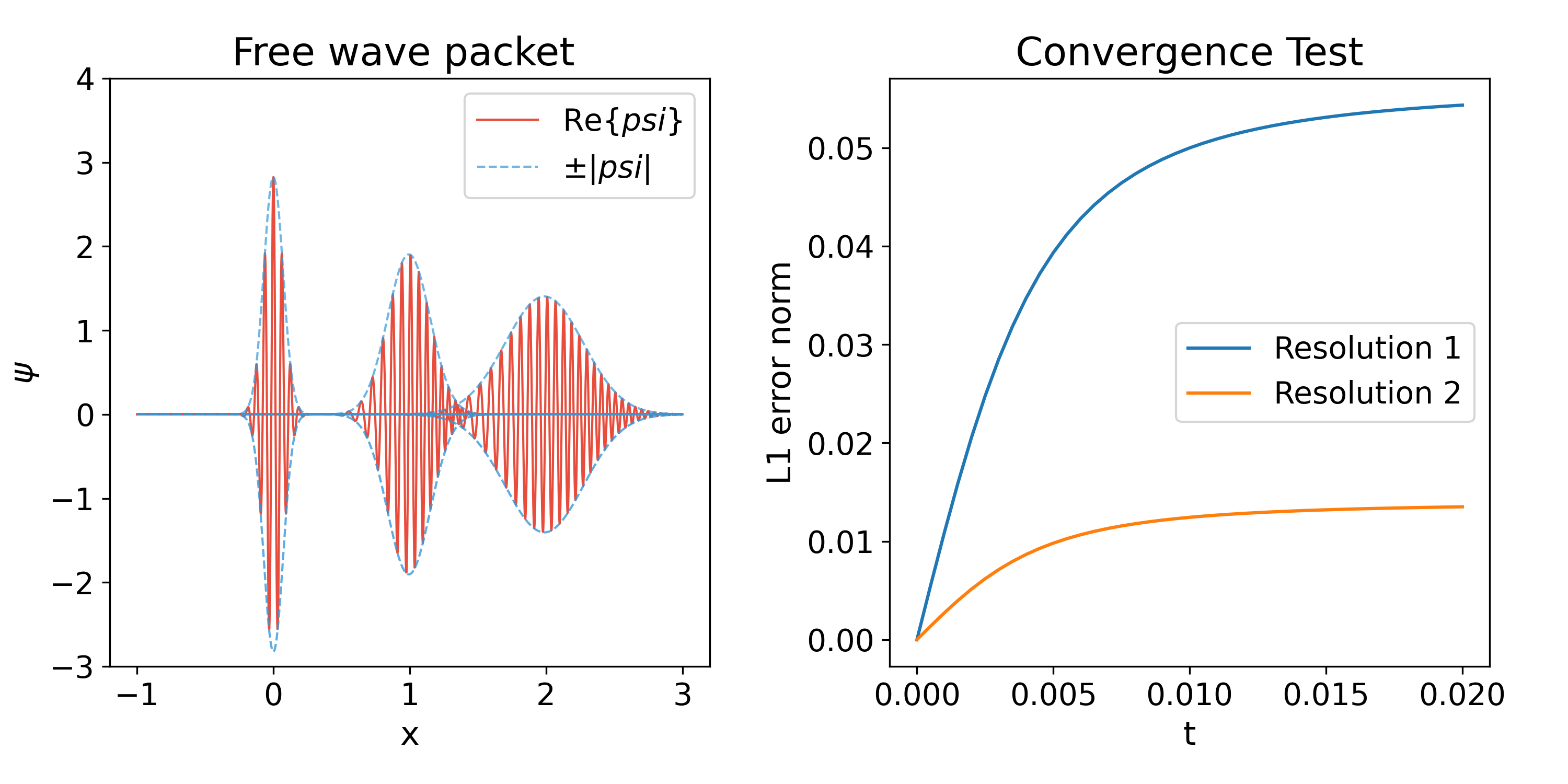}
\caption{At the left, snapshots of $\pm|\psi|$ and $\Re{\psi}$ at times $t=0,0.01,0.02$, showing how the wave packet spreads. At the right, convergence of the $L_1$ norm of the error to zero, of the density with respect to the exact solution (\ref{eq:ExactFreeParticle}).}
\label{fig:1dFreeParticle}
\end{figure}

\noindent What we do is evolve the initial conditions (\ref{eq:IDFreeParticle}) and illustrate the dynamics of the wave packet in Fig. \ref{fig:1dFreeParticle}. What happens is that the Gaussian density (\ref{eq:ExactFreeParticle}) widens and moves to the right. Fortunately there is an exact solution of this problem to compare with, so that we can show that the error between numerical and exact solution converges to zero with resolution.

This spreading property of particles reminds us of the Quantum Safari, defined by George Gamow in his book \cite{MrTompkins},\cite{MrTompkinsEnglish}.  When Mr. Tompkins and the professor go on an expedition to this place, they soon face a mosquito. But because of its quantum properties, after being spotted, its position becomes less certain over time. After a while, they are covered by a uniform mosquito probability density. This is exactly what happens with the Gaussian packet, $\Delta x$ grows indefinitely until it covers the entire space.

Another thing that can be done with the evolution is illustrate the difference between group and phase velocities. It is still difficult to explain the difference between the group and phase velocities in a few snapshot, thus we have placed an animation of this effect at \cite{GeneralSE}.

The numerical parameters used for the numerical domain are $x\in [ -1,3]$, $N_x = 2000$, $CFL = 0.125$, $N_t = 40000$, whereas the initial conditions in (\ref{eq:IDFreeParticle}) are $k_0 = 100$ and $a=0.1$. 

\subsection{1D Harmonic oscillator}

We now proceed to evolve a moving Gaussian wave packet subject to the effects of a harmonic oscillator potential. The potential and the initial conditions are:

\begin{eqnarray}
V(x) &=& \frac{1}{2}\omega^2x^2\nonumber\\
\psi(x,0) &=& e^{{\rm i} p_0 x} e^{-(x-x_0)^2/a^2}\label{eq:1dho}
\end{eqnarray}

\noindent where $p_0$ is the initial velocity of the wave packet of width $a$ initially centered at $x_0=0$. The classical analog would be a particle of mass $m$ attached to a spring, whose position and momentum are $x(t)$ and $p(t)$. These functions obey the following  equations of motion and initial conditions:

\begin{figure}
\centering
\includegraphics[width=9cm]{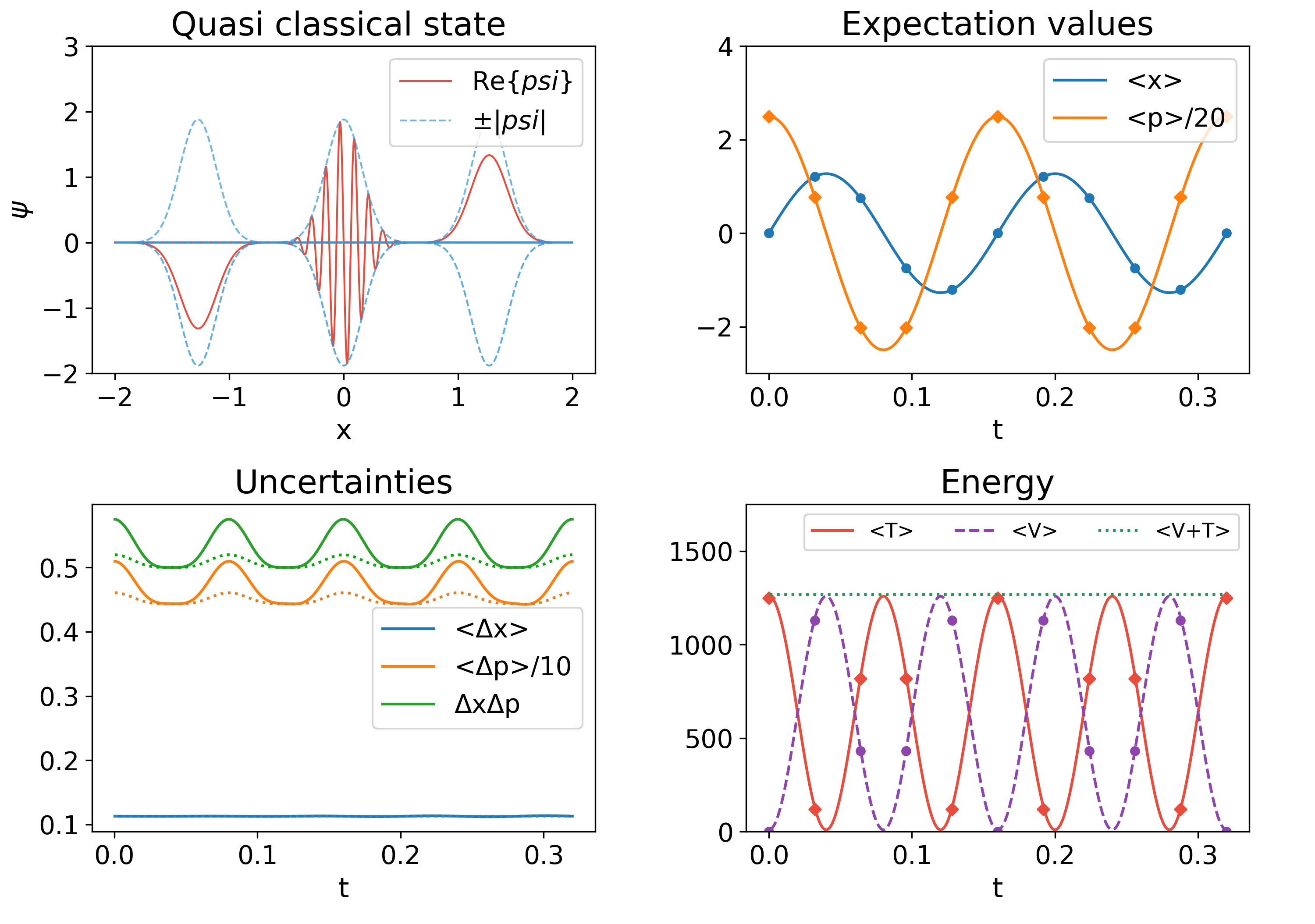}
\caption{Evolution of the quasi-classical wave packet onto a harmonic oscillator potential. At the top-left there are three representative snapshots, two at turning points and one at $x=0$, the minimum of the potential. At the top-right we draw the expectation values of the position and momentum as functions of time with continuous lines and the values of the equivalent classical oscillator with points. At the bottom-left we show the uncertainties in position and momentum along with their product for two resolutions, low/high with continuous/dashed lines that are different, which reveals the influence of numerical approximations; nevertheless, the extrapolation of the uncertainties obtained from  solutions with low and high resolutions is constant. Finally, at the bottom-right we show the behavior of the expectation values of the energy components as functions of time, where the points correspond to the values of the classical system at a few times.}
\label{fig:1dHO}
\end{figure}

\begin{eqnarray}
    \frac{dx}{dt} &=& p, \nonumber\\
    \frac{dp}{dt} &=& -\omega^2x,\nonumber\\
    x(0) &=&0, \nonumber\\
    p(0) &=& p_0,
\end{eqnarray}

\noindent whose solution reads:
\begin{eqnarray}
    x(t) &=& \frac{p_0}{\omega}\sin ( \omega t), \nonumber\\
    p(t) &=& p_0 \cos( \omega t).
    \label{eq:ClassicalSolution}
\end{eqnarray}

\noindent An interesting theorem that holds for all quantum mechanical systems is Ehrenfest's theorem; which probes the following equalities between expectation values:
\begin{eqnarray}
    \frac{d\langle \hat{x} \rangle}{dt} &=& \langle \hat{p} \rangle, \nonumber\\
    \frac{d\langle \hat{p} \rangle}{dt} &=& -\langle \nabla_x \hat{V} \rangle.
\label{eq:ehrenfest}
\end{eqnarray}

\noindent This set of equations looks very similar to Newton's second law, it seems like the expectation value of $\hat{x}$ follows the trajectory of a classical particle. Nevertheless, this is only true when $\langle \nabla \hat{V} \rangle = \nabla V (\langle \hat{x} \rangle)$ \cite{cohen}. In particular, this equality holds for the harmonic oscillator. Therefore, we can expect $\langle \hat{x} \rangle (t)$ and $\langle \hat{p} \rangle (t)$ to behave like $x(t)$ and $p(t)$ respectively if we start with $\langle \hat{x} \rangle (0) = 0$  and $\langle \hat{p} \rangle (0) = p_0$. 

This result is independent of the value of $a$. Nevertheless, when $a$ takes on the special value  $a = \sqrt{2/\omega}$, the total energy of the particle is minimum and the wave packet does not spread, a case called {\it quasi-classical state} \cite{cohen}. In Fig. \ref{fig:1dHO} we show snapshots of the evolution that illustrate how the pulse oscillates around the minimum. Two snapshots are taken at turning points and the other one at the minimum of the potential. Notice that at turning points the wave function has no nodes, whereas at the center the number of nodes is maximum, which is consistent with the fact that bigger the number of nodes the higher the kinetic energy. Also shown are the expectation values of position $\langle \hat{x} \rangle$ and momentum $\langle \hat{p} \rangle$ as functions of time, on top of them the points indicate the position and momentum of the classical system (\ref{eq:ClassicalSolution}) at some times. Turning points can be identified as the maximae and minimae of $\langle \hat{x} \rangle$ at times $0.08k+0.04$ with $k$ integer.

Additional information in Fig. \ref{fig:1dHO} are the uncertainties $\Delta x$ and $\Delta p$. As expected, since the wave packet does not spread, $\Delta x$ remains nearly constant whereas $\Delta p$ seems to change, however we verified that the variations are due to numerical errors, which is why we also show the uncertainties using a higher numerical resolution with dashed lines; in fact a Richardson extrapolation of the momentum uncertainty using the two resolutions  reveals that $\Delta p$ is constant. The product $\Delta x \Delta p$ is also illustrative. Notice that, in agreement with the uncertainty principle,  $\Delta x \Delta p$ is always $\ge 1/2$. Since the uncertainty in momentum is affected by numerical errors, then the product is also affected, reason why we also show in dashed line the product using high resolution. The extrapolation corresponds to a constant value in the continuum limit of $1/2$, which coincides with the theoretical value \cite{cohen}.

Finally, in the figure we also show the conservation of the expectation value of the total energy, whereas the potential $\langle T \rangle$ and kinetic $\langle V \rangle$ energies oscillate. Dots indicate the values of these energies in the classical case (\ref{eq:ClassicalSolution}) at various times.

\begin{figure}
\centering
\includegraphics[width=9cm]{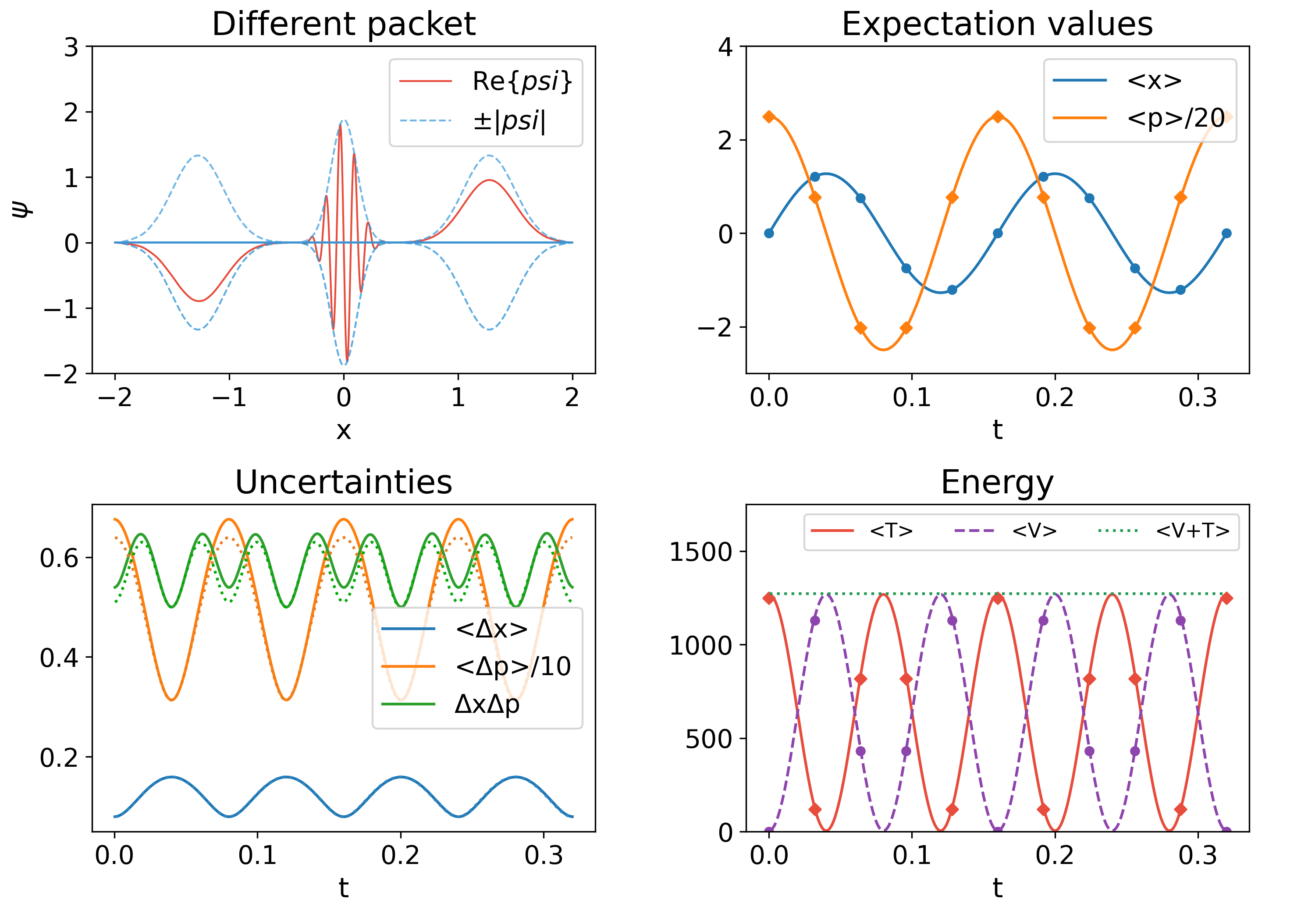}
\caption{Evolution of a wave packet different from the quasi-classical case, in a harmonic oscillator potential. At the top-left there are the three representative snapshots at center and turning points. At the top-right the expectation values of the position and momentum as functions of time are drawn with continuous lines and some values of the equivalent classical oscillator with points. At the bottom-left we show the uncertainties in position and momentum along with their product. Finally, at the bottom-right we show the behavior of the expectation values of the energy components as functions of time along with dots corresponding to the classical solution.}
\label{fig:1dHOB}
\end{figure}

In Fig. \ref{fig:1dHOB}, we present snapshots of the wave packet evolution with $a = \sqrt{1/\omega}$, different from the quasi-classical profile. It can be seen that the Gaussian spreads as it approaches the turning points. The expectation values of the position and momentum are still the same as the classical analogs, however the uncertainties look very different to the quasi-classical wave packet, for example in this case $\Delta x$ is not constant in time anymore but oscillates. Unlike in the quasi-particle wave packet, the uncertainties genuinely oscillate and an extrapolation of their values using higher resolution does not lead to constant values anymore. The figure shows that the uncertainty principle $\Delta x \Delta p \ge 1/2$ holds, as well as the conservation of total energy. An animation of the evolution of this system can be fount in \cite{GeneralSE}.

The parameters that define the numerical domain are the same for the two wave packets, namely 
$x\in [-2,2]$, $N_x = 2000$, $CFL = 0.125$, $N_t = 640000$, whereas the parameters for the initial conditions are $p_0 = 50$ and $\omega = 12.5\pi$.

\subsection{Perturbed harmonic oscillator}

A more dynamical scenario that involves a time dependent potential is the forced harmonic oscillator. For this, we perturb the ground state of the harmonic oscillator with a sinusoidal wave, which can correspond to a harmonic trap perturbed with an electromagnetic wave, a case used to illustrate resonance \cite{cohen}. The potential and initial condition used for this scenario are:

\begin{eqnarray}
V(x,t) &=& \frac{1}{2}\omega^2x^2+\alpha \omega^2 x \sin{\omega t}, \nonumber\\
\psi(x,0) &=& \left( \frac{\omega}{\pi} \right)^{1/4} e^{-\frac{1}{2}\omega^2x^2}.
\end{eqnarray}

\noindent Notice that the frequency of the perturbation coincides with the natural frequency of the harmonic potential, which is expected to produce resonance. For comparison we use the solution of the classical problem, given by:

\begin{equation}
    x(t) = \frac{\alpha}{2}\left( \omega t \cos{\omega t} - \sin{\omega t}\right),
\end{equation}

\noindent which is an oscillation of the particle with linearly growing amplitude.

\begin{figure}
\centering
\includegraphics[width=9cm]{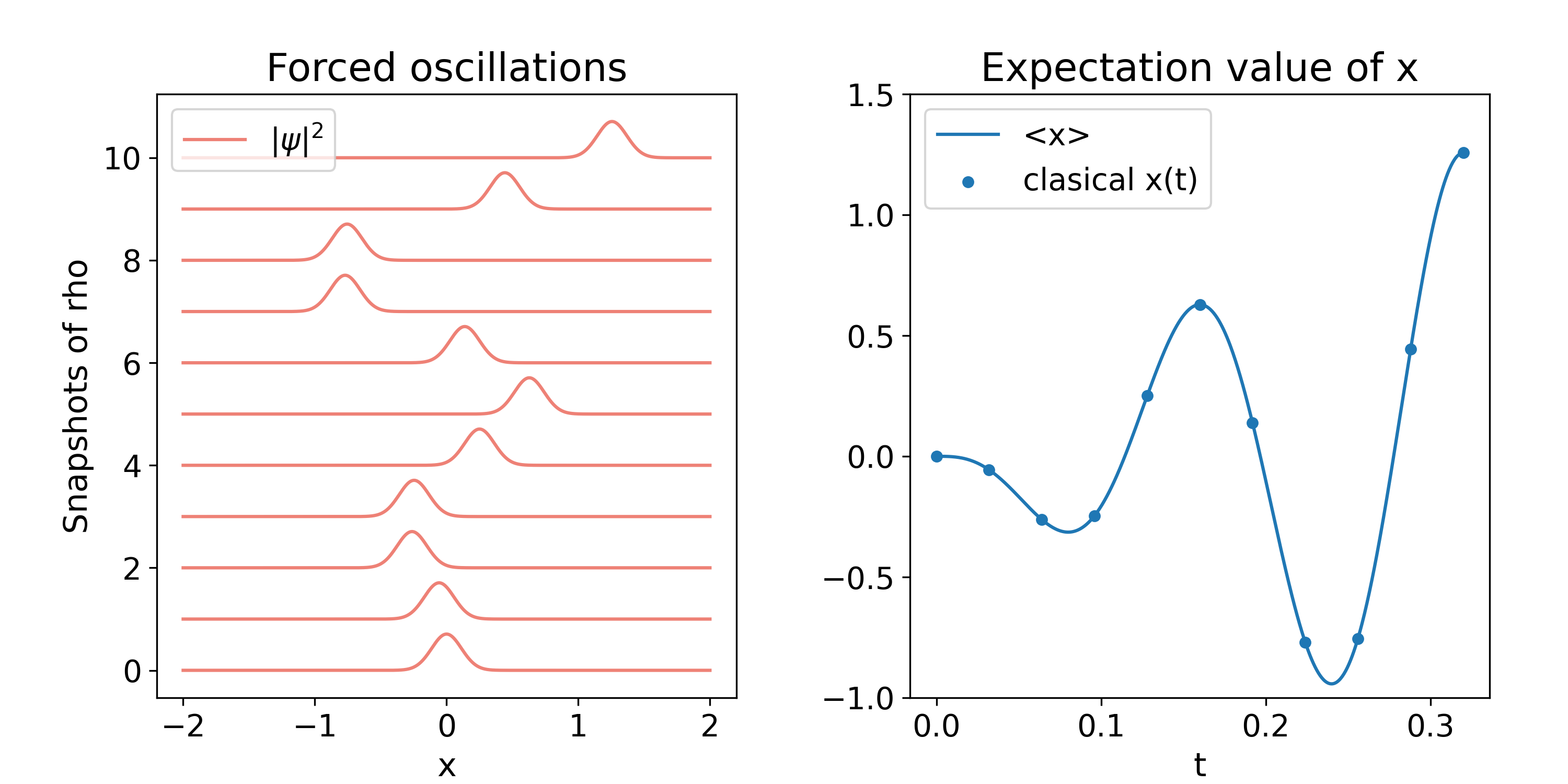}
\caption{On the left, snapshots of the probability density of the perturbed ground state taken every $t = 0.032$, earliest at the bottom and latest on top. On the right, the expectation value of $\hat{x}$ with continuous line and the classical solution $x(t)$ indicated with dots for a few values of time that coincide with those of the snapshots in the left graph.}
\label{fig:ForcedHO}
\end{figure}

We solve Schr\"odinger equation for the evolution of this system with $\alpha = 0.2$ and $\omega = 12.5 \pi$, on the numerical domain defined by $x\in [ -2,2]$, $N_x = 2000$, $CFL = 0.125$ and $N_t = 640000$. The results are shown in Fig. \ref{fig:ForcedHO}.
The wave packet oscillates with increasing amplitude, so that turning points are each time further out from the center. The amplitude growth reflects the resonance caused by the sinusoidal perturbation with the appropriate frequency $\omega$. The right graph of Fig. \ref{fig:ForcedHO} also shows that the Ehrenfest theorem continues to hold for this time dependent potential. Notice that the Gaussian packet preserves its shape during the evolution because the width used corresponds to the quasi-classical packet. For visual grasp we show the animation of this problem at \cite{GeneralSE}.

Finally, as far as we can tell, there is no exact solution to compare these results with.

\section{Problems in 2D}
\label{sec:p2d}

In this section we solve Schr\"odinger equation on two space dimensions that help illustrating the dynamics of wave-packets in widely used scenarios in basic Quantum Mechanics.

\subsection{2D Harmonic oscillator}

Our first example is the quasi-classical analog of the one-dimensional case above, which does not show dispersion of the density of probability. The potential and wave function profile we use are the following:

\begin{eqnarray}
V(x,y) &=& \frac{1}{2}\omega^2(x^2+y^2)\nonumber\\
\psi(x,y,0) &=&\sqrt{\frac{2}{\pi a^2}} e^{{\rm i} \vec{p_0}\cdot \vec{x}} e^{-((x-x_0)^2+(y-y_0)^2)/a^2},
\label{eq:2dho}
\end{eqnarray}

\noindent which corresponds to a circularly symmetric wave packet with momentum $\vec{p_o}$. The classical analog of this problem is a particle in the central potential: 

\begin{equation}
V(r) = \frac{1}{2}\omega^2r^2,
\end{equation}

\noindent where $r^2=x^2+y^2$. We also use the solution of the classical version of the problem and set initial conditions suitable for the particle's trajectory to be a circle. We do this by equating the centripetal acceleration to the force:

\begin{equation}
    \frac{p^2}{r} = \omega^2r ~~ \Rightarrow  ~~     p = \omega r,~~p = ||\vec{p}||.
\label{eq:2dclassicalA}
\end{equation}

\noindent We define initial conditions for the circular trajectory with $\omega = 12.5\pi$ and $r = 0.2$, then $\vec{p}$, with magnitude of $2.5\pi$, is chosen perpendicular to the position vector. The period of this trajectory would be:

\begin{equation}
    T = \frac{2\pi r}{p}=\frac{2\pi}{\omega} = 0.16,\label{eq:2dclassicalB}
\end{equation}

\noindent which is independent of the radius of the trajectory.

\begin{figure}
\centering
\includegraphics[width=9cm]{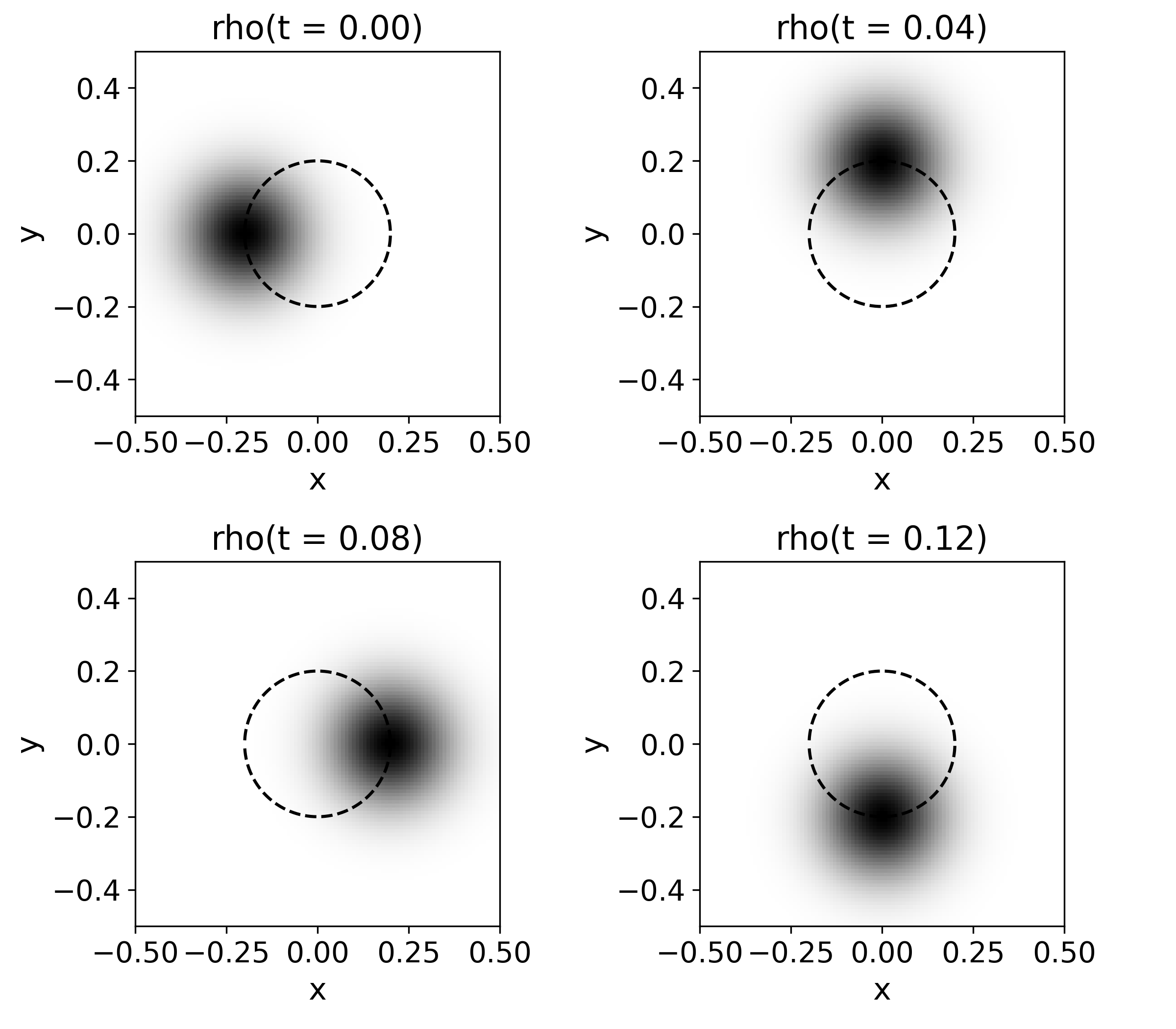}
\caption{Snapshots of the evolution for the wave packet corresponding to the quasi-classical solution on a two-dimensional harmonic oscillator travelling on a circular trajectory. The initial position and momentum are $-0.2\hat{\textbf{\i}}$ and $0.2\omega \hat{\textbf{\j}}$ respectively. The dotted circle is the trajectory formed by the expectation value of the position $\langle \hat{\vec{x}} \rangle$, which describes a circle of radius $0.2$. In fact, the solution of the classical problem would be exactly on top of this circle as well. The shape of the wave packet does not disperse away while it orbits around the origin.}
\label{fig:2DquasiClasic}
\end{figure}

{\it Quasi-classical case.} In Fig. \ref{fig:2DquasiClasic} we illustrate the result of the evolution of this wave packet. The wave packet indeed moves on a circle of radius $0.2$ with a period of $0.16$ in the clockwise direction. The shape of the packet does not change because the width of the packet is set to the special value $a = \sqrt{2/\omega}$. This width is that of the ground state of the two-dimensional harmonic potential. For these reasons, we call this packet the {\it quasi-classical solution}.
The circle in this Figure corresponds to the trajectory of a classical particle with the same initial conditions under the influence of the classical potential described by Eqs. (\ref{eq:2dclassicalA}) and (\ref{eq:2dclassicalB}), and also corresponds to the trayectory of the expectation value of the position $\langle \vec{x}\rangle (t)$. This solution of Schr\"odinger equation was calculated in the numerical domain defined by $x \in [-1,1]$, $y \in [-1,1]$, $Nx = Ny = 200$, $CFL = 0.125$ and $Nt = 12800$, which corresponds to $t_f = T$

\begin{figure}
\centering
\includegraphics[width=9cm]{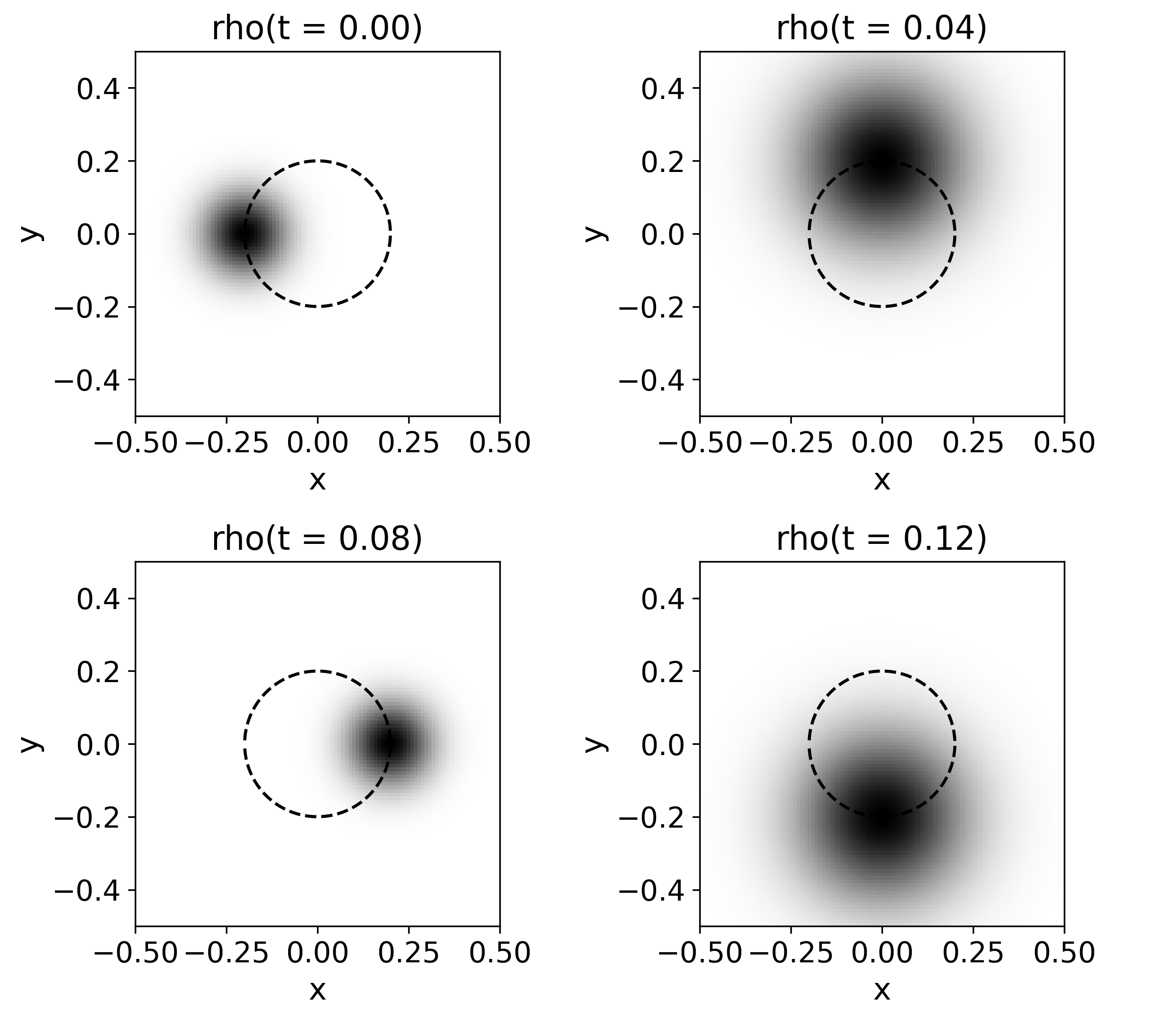}
\caption{Evolution of the wave packet on the two-dimensional harmonic oscillator potential with $a = \sqrt{1/\omega}$. The initial position and momentum are the same as for the quasi-classical state. The expectation value of the position follows the same trajectory as the classical analog likewise in the quasi-classical particle case. The wave packet widens when it crosses the $y-$axis and recovers its initial shape when it crosses the $x-$axis during the evolution.}
\label{fig:2DnonTrivial}
\end{figure}

{\it Another case.} Likewise in the 1D scenario we also present the case with a non-educated Gaussian pulse with $a = \sqrt{1/\omega}$. The evolution is very interesting and the density appears in Fig. \ref{fig:2DnonTrivial}. The pulse evolves in clockwise direction. It starts with a concentrated probability distribution and like in the one dimensional case, it starts spreading as can be seen in the second snapshot, until it crosses the $y-$axis; afterwards, the packet begins to compress again until it acquires its original shape at the $x-$axis again in the third snapshot. The evolution repeats itself with the pulse compressing at the $x-$axis and most expanded at the $y-$axis. 

The circle in this Figure corresponds to the expectation value $\langle \vec{x}\rangle (t)$ as well as the trajectory of the classical particle with the same initial conditions under the influence of the classical potential. One takeaway from this exercise is that the Ehrenfest theorem still holds, that is, the expectation value of the position keeps behaving like that of the equivalent classical particle when we jump to two dimensions. 

An animation of the quasi-classical state and one with crazy initial conditions can be seen at \cite{GeneralSE}.

\subsection{Diffraction by a single slit }

This problem is defined on a two space dimensions domain as follows. A Gaussian wave packet is launched towards a potential wall with a ``hole''. The wave function partly bounces from the potential wall and partly passes through. What is illustrative is the well known diffraction pattern that has to be detected beyond the wall.

The initial condition for the wave function is $\psi(x,y,0)=\sqrt{\frac{2}{\pi a^2}} e^{{\rm i} p_0x} e^{-((x-x_0)^2+y^2)/a^2}$, which has a Gaussian profile with a momentum along the $x-$direction. The potential wall is centered at the $y-axis$ and technically has a finite thickness of five domain cells. The slit is implemented as a segment of the $y-$axis where the potential is zero. This potential can be expressed as follows:

\begin{equation}
V(x,y) = \left\{
\begin{array}{ll}
    V_0  & (x,y)\in [-0.025,0.025]\times \nonumber\\
    & ( [-1,-0.15) \cup (0.15,1])\nonumber\\
    & \nonumber\\
    0 & {\rm else}\nonumber
\end{array}\right.
\end{equation}

\noindent The potential barrier is high enough, with $V_0 = 4000$, as to mimic an infinite  potential. Technically this avoids one the need to implement zero boundary conditions on the wave function at the boundary of the wall.
The domain parameters used for the numerical solution are $x \in [-1,1]$, $y \in [-1,1]$, $N_x = N_y = 200$, $CFL = 0.125$ and $Nt = 1950$, whereas the initial conditions are defined by $x_0 = -0.4$, $p_0 = 12.5\pi$, $a = 0.25$.

\begin{figure}
\centering
\includegraphics[width=9cm]{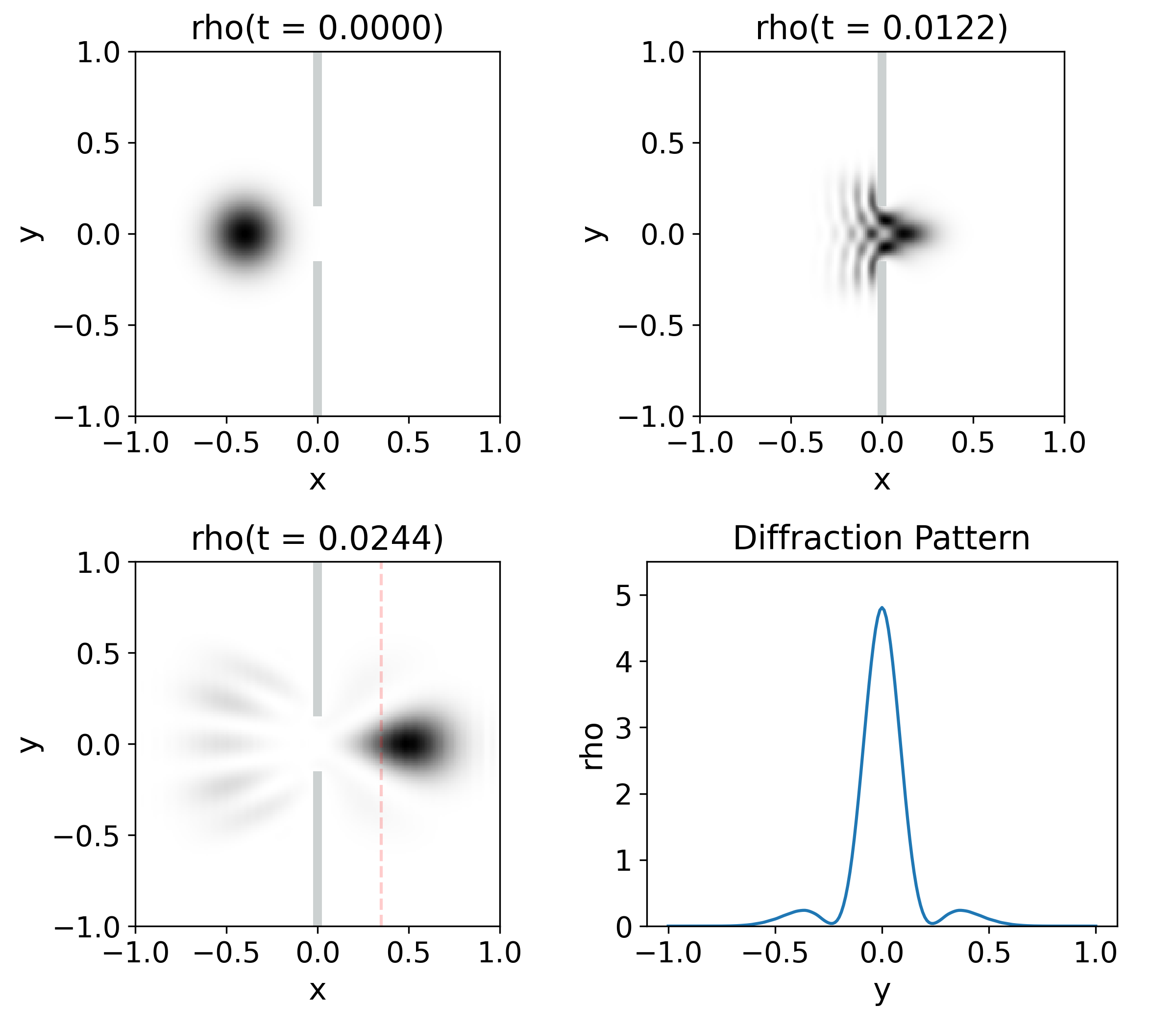}
\caption{Three snapshots of the density at different times prior and post interaction with the slit of width $0.3$. In the third snapshot we show the location of a detector at the line $x=0.35$, where we measure the density at time $t=0.0244$ and show the resulting diffraction pattern in the fourth plot. This pattern resembles part of the sinc function, expected for monochromatic light. }
\label{fig:diffraction}
\end{figure}

In Fig. \ref{fig:diffraction}, one can see the evolution of the wave packet during and after passing through the slit. When the packet arrives at the slit, it starts interfering with itself and then a major part of the wave passes through. The reflected pulse looks interesting, with various fringes although not very well discussed in books. Concerning the pulse that passes through, the pattern forward is the one most analyzed. In order to picture this pattern, we implemented a detector screen placed at $x = 0.35$, where we measure the probability density, shown on the fourth image of Fig. \ref{fig:diffraction}, measured at the time of the third snapshot. The pattern is composed of three intensity peaks and two minima located at $y = \pm 0.22$. This result can be compared with the Fraunhofer diffraction of a wave going through a slit, which is $\sin(u)/u$ function with minima located at the following positions \cite{hecht}:

\begin{equation}
    a\sin\theta = m\lambda,~~m = \pm 1, \pm 2,...
\end{equation}

\noindent where $a$ is the width of the slit, $\lambda$ the wave length, $\theta$ the deflection output angle and $m$ the label of the minima \cite{hecht}. If we substitute the values from our simulation, we find that the first minima should be located at:

\begin{equation}
\sin \theta = \frac{\lambda}{a} = \frac{2\pi/p_0}{0.3} = \frac{8}{15},
\end{equation}

\noindent which, in the case of a screen located $0.35$ units of length away from the slit, corresponds to a $y$ value of $0.2206$. This corresponds to an error of $0.3\%$ which can be attributed to factors like the non monochromatic nature of the wave packet, and the fact that the distance between the slit and the screen is too small to be considered in the Fraunhofer regime. The important result is that, before the slit, the particle was not completely localized, but we knew that its momentum was around a specific value with $\Delta p = {\rm contant}$. However, after the slit, the particle could be directed towards any of the diffracted rays, and there is no {\it a priori} clue of which path it might have taken until we measure it. One interesting fact is that we can be sure that it will not be detected at angles like $\pm 32{}^{o}$ because of its wave-like properties.

\subsection{Experiment of the double slit}

One of the most used examples of Quantum Mechanics is the experiment of the double slit. A single particle governed by Schr\"odinger equation is thrown against a barrier with two slits. After the barrier, a screen detector is placed to measure the position of the particle. The experiment is performed multiple times and a detecting plate records the distribution of arriving positions. If the particle were to behave classically, one should only see two spots in the screen, each one associated to the pass of the particle through each of the slits. Nevertheless, what is observed is a set of stripes that reassembles an interference pattern produced by a wave. This is what we will reproduce.

The particle state will be represented by the wave function $\psi$ on a two-dimensional space. The barrier with two slits will be represented by a potential barrier of height $V_0$ and $0.05$ units thick. The two slits are separated by the distance $d = 0.4$ and $0.1$ units wide. The explicit potential is:

\begin{equation}
V(x,y) = \left\{
\begin{array}{ll}
    V_0  & (x,y)\in [-0.025,0.025]\times \nonumber\\
    & ( [-1,-0.25) \cup (-0.15,-0.15) \cup (0.25,1])\nonumber\\
    & \nonumber\\
    0 & {\rm else}\nonumber
\end{array}\right.
\end{equation}

\noindent The initial value of $\psi$ will be that of a Gaussian wave packet: $\psi(x,y,0)=\sqrt{\frac{2}{\pi a^2}} e^{{\rm i} p_0x} e^{-((x-x_0)^2+y^2)/a^2}$ with $a = 0.3$, $x_0 = -0.4$, $p_0 = 12.5\pi$ and $V_0 = 6000$. Finally, the domain parameters are the following: $x \in [-1.5,1.5]$, $y \in [-1.5,1.5]$, $Nx = Ny = 300$, $CFL = 0.125$ and $Nt = 2000$.

\begin{figure}
\centering
\includegraphics[width=9cm]{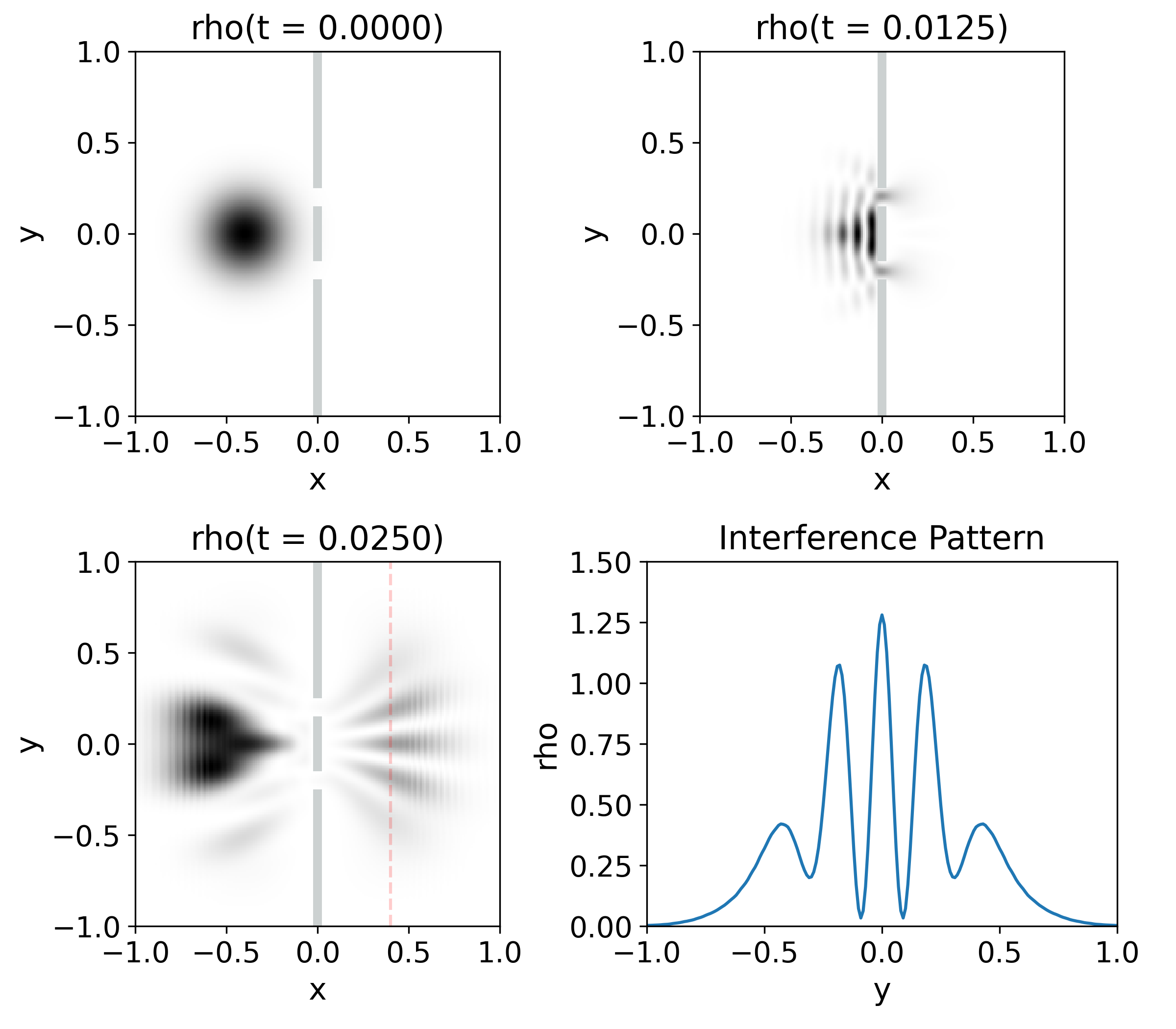}
\caption{Snapshots of the evolution of the Gaussian packet initially moving to the right and colliding against the potential barrier with two slits. Part of the packet is reflected and part of it goes through, creating an interference pattern. On the third snapshot, a screen detector is placed along the line $x = 0.4$. The density measured along such line is shown on the fourth figure that records the interference pattern, which reassembles that of Young interferometer. }
\label{fig:interference}
\end{figure}

Figure \ref{fig:interference} illustrates the behavior during the evolution of the wave packet. After the barrier, the expected interference pattern is formed. The location of the intensity peaks can be compared with the ones predicted by the Young interferometer, originally designed for electromagnetic waves \cite{hecht}. According to Young, the intensity peaks are located at the angles given by the formula 

\begin{equation}
d\sin\theta = m\lambda,~~ m = 0,\pm 1, \pm2,....
\end{equation}

\noindent In our case, $d = 0.4$ and $\lambda = 0.16$, thus the zero-th order peak should be at $y=0$, the first order peak should be located at $\theta=\pm 23.57{}^{o}$  and the second order peak at $\theta = \pm 53.13{}^{o}$. We measure the density at a screen located at the line $x=0.4$, where  the first and second order peaks would be centered at $y = \pm 0.1745$ and $y=\pm 0.5333$ respectively. From our simulations these peaks are centered at $y=\pm 0.18$ and $y=\pm 0.43$. The error between the prediction and our measurements are $3.1\%$ and $19.32\%$ respectively, that we attribute to both, the approximations of Young formula and the non-monochromatic nature of the Gaussian packet. 

The appealing interference pattern seen in the third snapshot of Fig. \ref{fig:interference} reminds again of the Quantum Safari \cite{MrTompkins},\cite{MrTompkinsEnglish}. In one of the scenes described, a flock of gazelles grazed peacefully when a lioness suddenly appeared. The gazelles ran scared against a row of trees with two small gaps. After crossing the trees, the gazelles divided in columns each one headed directly against a group of hungry lionesses. The gazelles were actually behaving like the quantum particle of Fig. \ref{fig:interference}, and the lionesses where position in the intensity peaks of the interference pattern.


Animations of the single and double slit problems are found at \cite{GeneralSE}.

\subsection{Reflection and Refraction}

In this example we illustrate what happens when a wave packet collides against a finite height step potential in two dimensions:

\begin{equation}
V(x,y) = \left\{
\begin{array}{ll}
    0  & x \leq 0,~y \in D\\
    & \\
    V_0 & x > 0, ~y\in D \label{eq:2dsteppotential}
\end{array}\right. .
\end{equation}

\noindent This setting is just like the case when light changes from one medium to another, in the quantum case, particle-waves get reflected and refracted when they face a step potential. For a plane monochromatic wave, the laws that rule this phenomenon are the reflection and refraction laws:

\begin{eqnarray}
    \theta_i &=& \theta_r ,\nonumber\\
    k\sin \theta_i &=& \sqrt{k^2-k_0^2}\sin \theta_t,
\label{eq:snellModificada}
\end{eqnarray}

\noindent where $\theta_i$, $\theta_r$ and $\theta_t$ are the angles of incidence, reflection and transmission, considering the same conventions used in optics \cite{hecht}; $k$ is the wave number and $k_0^2 = 2mV_0/\hbar^2$ or $2V_0$ in code units. One very interesting fact is that a {\bf classical particle} that goes through the same potential follows the refraction law when its horizontal velocity is greater or equal to $\sqrt{2V_0}$; and follows the reflection law when the horizontal velocity is not big enough.

\begin{figure}
\centering
\includegraphics[width=8.5cm]{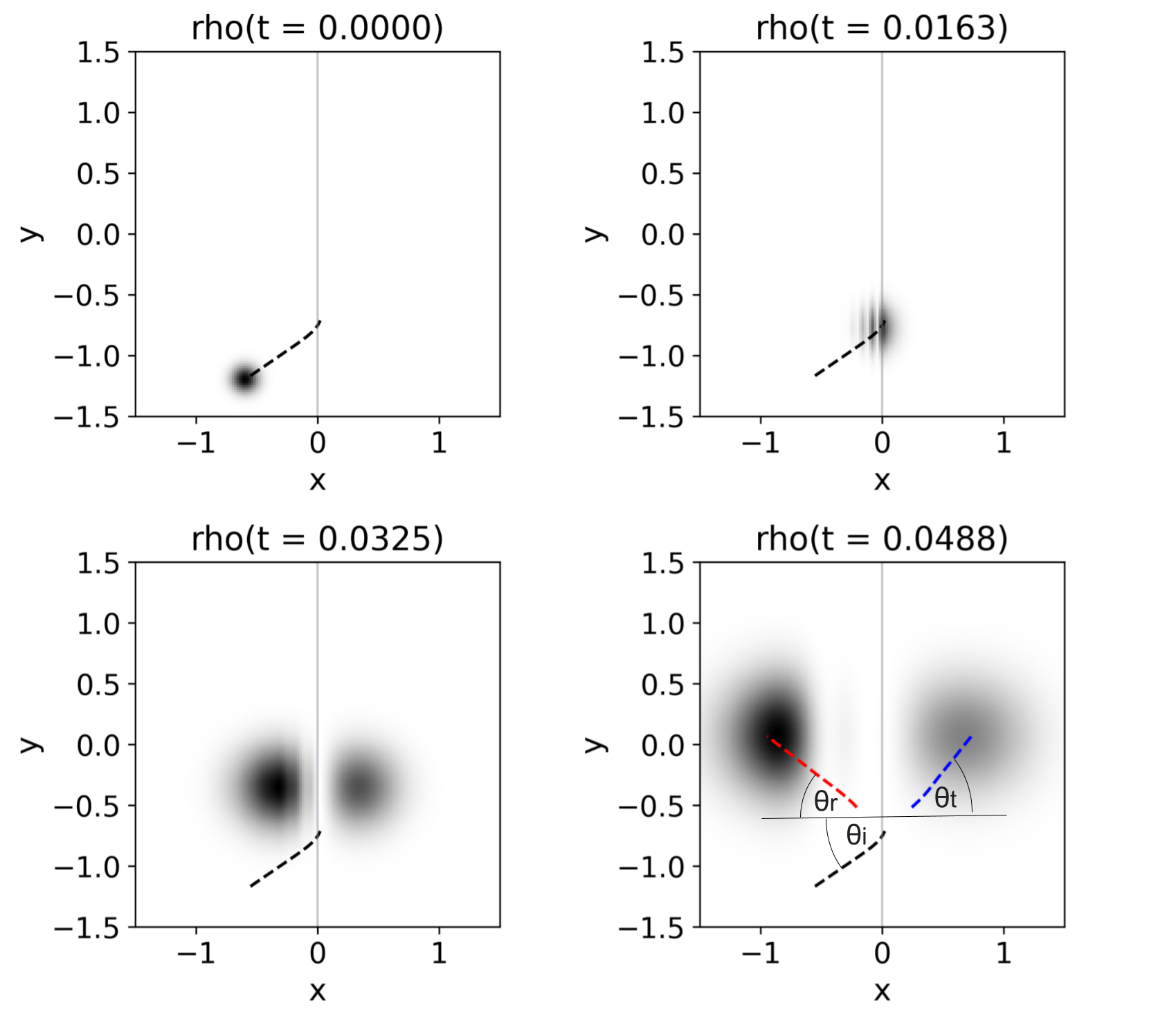}
\caption{Evolution of a Gaussian wave packet moving with oblique incidence against the step potential (\ref{eq:2dsteppotential}). In this case, $k = 15\pi$, $V_0 = 750$ and $\theta_i = 34{}^{o}$. The numerically calculated reflection and transmission angles are: $\theta_{r} = 38.93{}^{o}$ and $\theta_t = 52.13{}^{o}$. The dashed paths are the trajectories followed by the expectation value $\langle \vec{x}\rangle (t)$ of the initial and the two resulting pulses, reason why they are not quite straight lines.}
\label{fig:refraction}
\end{figure}

In Fig. \ref{fig:refraction} appears the density of a Gaussian wave packet initially located at $(-0.6,-1.2)$ with momentum $||\vec{p_0}|| = 15\pi$ heading towards the potential step with an incidence angle of $34{}^{o}$. In the second snapshot, the wave packet hits the interface and interferes with itself. In the third snapshot the packet splits into two as the result of the interaction with the step: one bump reflected and another one refracted.  Notice that the black dashed line indicates the trajectory of $\langle \vec{x} \rangle (t)$ of the packet prior to the collision. Finally, in the last snapshot, we indicate with red and blue dashed lines the trajectory that $\langle \vec{x} \rangle (t)$ follows for each of the two wave packets. From these expectation values, we find that the reflection and transmission angles are approximately $\theta_r = 38.93º$ and $\theta_t = 52.13º$ respectively. A classical particle, as well as a monochromatic wave, would refract with an angle of $80{}^{o}$. We can see that the value obtained numerically doesn't match the theoretical one very well. This is because the refraction angle depends on the wave number, not only on $\theta_i$. Since the wave packet is a sum of plane waves with different momentum, each of the plane waves refracts at a different angle, changing the shape of the packet and the refraction angle of $\langle \vec{x} \rangle$.

The domain parameters used for this simulations are $x \in [-2,2]$, $y \in [-2,2]$, $Nx = Ny = 400$, $CFL = 0.125$ and $Nt = 3900$, and the initial conditions are set to $\vec{x}_0 = -0.6\hat{\textbf{\i}}-1.2\hat{\textbf{\j}}$, $\vec{p}_0 = 15\pi (\cos (34{}^{o})\hat{\textbf{\i}}+\sin(34^{o}) \hat{\textbf{\j}})$, packet width $a = 0.15$ and $V_0 = 750$.

siddh
\subsection{Dispersion by a central potential}

As a final example, we look at a case of dispersion by a central potential. In this experiment, a particle (or wave packet) is directed against a potential that is spherically symmetric. One not so obvious assumption that is made by \cite{cohen} for the treatment of this problem, is that after the plane wave interacts with the potential, the wave function becomes the sum of a plane wave that goes through and a spherical wave front with nonuniform amplitude. The objective of this section is to show numerically that the outgoing wave function truly has a spherical component. 

We threat the two dimensional case, with potential and initial conditions as follows:

\begin{eqnarray}
V(x,y) &=& V_0e^{-(x^2+y^2)/2\sigma^2},\nonumber\\
\psi(x,y,0) &=&\sqrt{\frac{2}{\pi a^2}} e^{{\rm i} p_0x} e^{-((x-x_0)^2+y^2)/a^2}.
\label{eq:centralPotential}
\end{eqnarray}

\begin{figure}
\centering
\includegraphics[width=8.5cm]{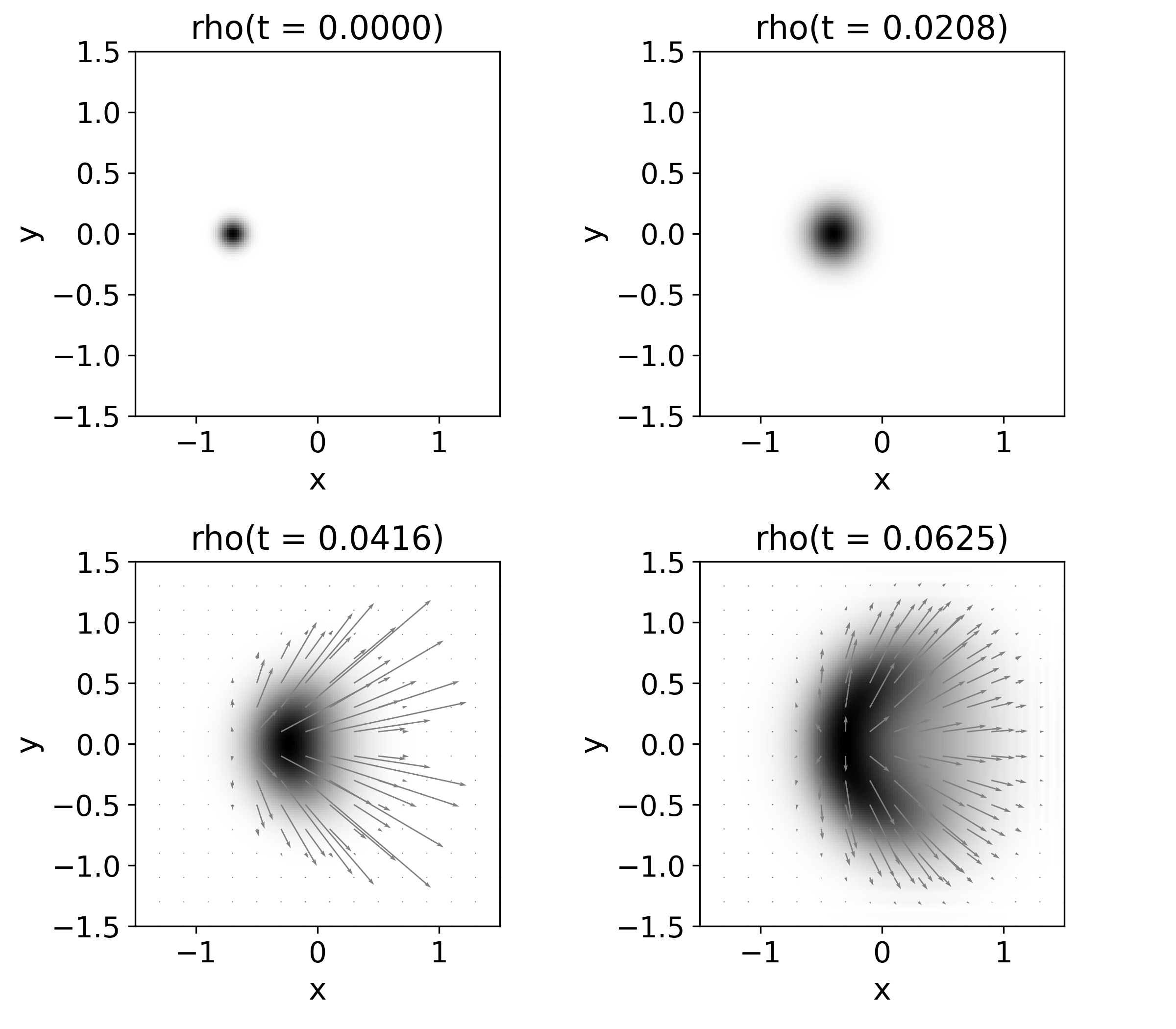}
\caption{Evolution of a Gaussian wave packet directed against the origin of the central potential of equation (\ref{eq:centralPotential}). In this case $p_0=5\pi$, $V_0 = 120$ and $a = 0.15$. After the interaction with the potential, the wave function appears to have a circular wave front. In the last two snapshots, the black spot grows in size in all directions. A map of the probability current vector field is on top. The current emerges radially from the center, suggesting a circular but nonuniform wavefront.}
\label{fig:centralPotential}
\end{figure}

\noindent In Fig. \ref{fig:centralPotential}, one can see the outcome of the interaction between the wave packet and the central potential, which in this case is repulsive. The packet starts moving to the right, but as it approaches the origin it slows down and deforms to avoid the potential peak. The particle then becomes a blur that starts spreading in all directions. In the last two snapshots, the probability current is mapped on top of $\rho$. The vector field looks almost radial, evidence of the particle spreading in all directions as predicted \cite{cohen}. 

The domain parameters used for this simulations were: $x \in [-1.5,1.5]$, $y \in [-1.5,1.5]$, $Nx = Ny = 300$, $CFL = 0.125$ and $Nt = 3900$, and initial conditions $x_0 = -0.7$, $p_0 = 5\pi$, $a = 0.15$, $V_0 = 120$ and $\sigma = 0.3$.


\section{Final comments}
\label{sec:comments}

We have solved the time dependent Schr\"odinger equation using numerical methods in scenarios involved with the wake-particle duality.

We expect these examples help illustrating the robustness of Schr\"odinger equation solutions in the wave and particle interpretation of elementary Quantum Mechanics. 
More detailed pictures and animations corresponding to the examples described in this paper are available at \cite{GeneralSE}, which can be useful for teaching purposes. Finally, we want to stress that also for educational purposes, the codes needed for the reproduction of these results can be available under request.


\section*{Acknowledgments}
This work is supported by grant CIC-UMSNH-4.9.


\bibliography{QC}

\end{document}